\documentclass[a4paper,groupcitations]{article}
\usepackage{mwe}
\usepackage[T1]{fontenc}
\usepackage[ansinew]{inputenc}
\usepackage[english]{babel}
\usepackage{amsfonts}
\usepackage{amsmath}
\usepackage{array}
\usepackage{amsthm}
\usepackage{amssymb}
\usepackage{graphicx}
\usepackage{subfigure}
\usepackage{braket}
\usepackage{eucal}
\usepackage{verbatim}
\usepackage[table]{xcolor}
\usepackage{caption}
\usepackage{cite}
\usepackage{textcomp}
\usepackage{tikz}
\usetikzlibrary{positioning,arrows}
\usetikzlibrary{decorations.pathmorphing}
\usetikzlibrary{decorations.markings}
\usetikzlibrary{calc,decorations.markings}
\usetikzlibrary{arrows,shapes}
\usetikzlibrary{matrix,arrows}
\usepackage{pgfplots}
\usepackage{xparse}
\definecolor{jade}{HTML}{00A86B}
\newcommand{\be}{\begin{eqnarray}}
\newcommand{\ee}{\end{eqnarray}}

\newcommand{\expec}[1]{\mbox{$\langle\, #1\,\rangle$}}

\newcommand{\lpl}{\ell_{\rm p}}
\newcommand{\mpl}{m_{\rm p}}
\newcommand{\gn}{G_{\rm N}}

\title{\bf Bounded compactness from G(E)UP}
\author{Roberto~Casadio$^{ab}$\thanks{E-mail: casadio@bo.infn.it},
$\ $
Ilim Irfan \c{C}imdiker$^{c}$\thanks{E-mail: ilim.cimdiker@phd.usz.edu.pl},
$\ $
Octavian~Micu$^d$\thanks{E-mail: octavian.micu@spacescience.ro}
$\ $
and
Jonas~Mureika$^{e}$\thanks{Email: jmureika@lmu.edu}
\\
\\
$^a${\em Dipartimento di Fisica e Astronomia, Universit\`a di Bologna}
\\
{\em via Irnerio~46, 40126 Bologna, Italy}
\\
\\
$^b${\em I.N.F.N., Sezione di Bologna, I.S.~FLAG}
\\
{\em viale B.~Pichat~6/2, 40127 Bologna, Italy}
\\
\\
$^c${\em Institute of Physics, University of Szczecin}
\\
{\em Wielkopolska 15, 70-451 Szczecin, Poland}
\\
\\
{\em $^d$Institute of Space Science, Bucharest, Romania}
\\
{\em P.O. Box MG-23, RO-077125 Bucharest-Magurele, Romania}
\\
\\
{\em $^e$Department of Physics, Loyola Marymount University}
\\
{\em Los Angeles, California, USA}
}
\begin{document}
\maketitle
%
%
\begin{abstract}
We analyse how different Generalised Uncertainty Principles could place
bounds on the compactness of self-gravitating systems.
By considering existing experimental bounds on the relevant parameters,
we conclude that the compactness of large astrophysical objects is bounded
above by the inverse of the GUP parameter, which would naturally be of order one.
Conversely, the existence of black holes imposes stronger bounds on those
parameters.
\end{abstract}
\section{Uncertainty Principles and Quantum Gravity}
\setcounter{equation}{0}
\label{S:intro}
The nature of gravity and quantum physics seem to imply that physical measurements cannot probe arbitrarily
short length scales~\cite{Hossenfelder:2012jw,Casadio:2022opg}.
Typically, a minimum scale is assumed of the order of the Planck length on dimensional grounds and
considering that black hole geometries are supposed to emerge in scattering processes at Planckian
energies~\cite{Scardigli:1999jh}.
However, semiclassical studies of collapsing compact objects (possibly of astrophysical size), generically
predict a bounce at a scale that remains largely undetermined~\cite{Casadio:1998yr}, and corresponds
to a quantum transition from the formation of a black hole to a white hole geometry~\cite{Haggard:2014rza}.
Moreover, a quantum description of the (very large number of) particles forming such objects 
shows that the ground state at the bounce could be a (significant) fraction of the horizon radius,
hence many orders of magnitude larger than the Planck length~\cite{Casadio:2023ymt}.
These results suggest that quantum physics might not just imply the existence of a minimum length
but also of a maximum compactness for self-gravitating objects~\cite{Casadio:2021cbv}.
\par
A rather straightforward way of implementing a minimum length in quantum mechanics is to
consider a Generalised Uncertainty Principle (GUP) for the position $x$ and momentum $p$
of a particle, which can be written as
\be
\Delta x\, \Delta p
\geq
\hbar
\left(
1
+
\alpha\,\frac{\Delta p^2}{\mpl^2}
\right)
\ ,
\label{gup0}
\ee
where the dimensionless parameter $\alpha$ denotes the additional uncertainty introduced
by the formalism and $\mpl$ is the Planck mass.
The minimum length is then given by $\ell_{\rm min} \sim \sqrt{\alpha}\,\lpl$, where $\lpl$ is the
Planck length.
The literature is replete with studies devoted to the GUP, mostly as they relate to the quantum
gravity regime.
The GUP itself is a common feature of a variety of quantum gravity theories, including 
string theory~\cite{veneziano_1,veneziano_2,Scardigli:1999jh,veneziano_4,veneziano_5,veneziano_6},
loop quantum gravity~\cite{ashtekar_1,ashtekar_2}, ultraviolet self-complete gravity~\cite{nicolini},
non-commutative quantum mechanics~\cite{majid,Kanazawa:2019llj}, and generic minimum length
models~\cite{maggiore_1,maggiore_2,maggiore_3}.
A full summary of GUP black hole related works would be exhaustive, but the interested reader
is referred to~\cite{Adler_2,Adler_3,gup1,gup2,gup3,mijmpn,gupextrad,bcjmpn} and references therein.
A recent work~\cite{Carr:2024brs} provided a comprehensive study of quantum and classical constraints
on the GUP parameter as it relates to the ``$M+1/M$'' model presented in~\cite{bcjmpn}.
Such GUP effects are strictly relegated to the domain of Planck-scale black holes and their thermodynamics,
so this work will provide a new perspective on the GUP as it relates to astrophysical scale black holes. 
\par
Compared to the GUP, the literature on the Extended Uncertainty Principle (EUP) is sparse,
but has been growing steadily over the past few years.
As a complement to the GUP, the EUP introduces a minimum momentum scale, or alternatively
a maximum length scale, at which quantum gravity effects will manifest themselves.
This is written as
\begin{equation}
\Delta x \,\Delta p
\geq
\hbar
\left(1
+
\frac{\bar{\beta}}{L_{*}^2}\, \Delta x^2
\right)
\ ,
\label{eup0}
\end{equation}
where $L_*$ is some fundamental large length, possibly, but not necessarily, of the Hubble
scale.
In fact, the EUP implies a minimum momentum, corresponding to a maximum length scale.
A number of studies have therefore considered large scale EUP effects on black
holes~\cite{Bambi:2007ty,eup1,eup2,Cadoni:2018dnd,Mureika_2019}.
The weak field limit of an EUP-modified Schwarzschild black hole metric was also derived
in~\cite{Mureika_2019}, where it was suggested that the length scale $L_*$ can fix the flattening
radius of galactic rotation curves, and thus provide an alternative to dark matter.
Additional related studies include the effect of the EUP on strong lensing effects in Sgr~A*
and M87~\cite{luxie}, the Shapiro time delay, Mercury's perihelion shift, and S2 star orbit
precessions~\cite{Okcu:2022sio}.
Ref.~\cite{Illuminati:2021wfq} also explored lensing effects, in addition to solar spin precession
and the motion of binary pulsar systems.
Lastly, the EUP's influence on cosmology has been addressed, namely through connections
with the Hubble tension~\cite{Nozari:2024wir} and the cosmological constant been
addressed~\cite{Pantig:2024asu}.
Most recently, the authors of~\cite{Capozziello:2025iwn} have shown the EUP to be emergent
from non-local gravity effects.
\par
The combination of the GUP and the EUP is known as the Generalised Extended Uncertainty
Principle (GEUP), and assumes the form
\begin{equation}
\Delta x \,\Delta p
\geq
\hbar
\left(
1+\alpha\,\Delta p^2 +  \frac{\bar{\beta}}{L_{*}^2}\, \Delta x^2
\right)
\ ,
\label{geup0}
\end{equation}
The main aim of the present work is to see if the existence of a minimum length of the order
of the Planck scale generically implies bounds on the compactness of self-gravitating
systems. 
To investigate this question, we will consider the GUP~\cite{Scardigli:1999jh}
and the GEUP~\cite{Bambi:2007ty} as effective ways to incorporate
the existence of a minimum length in quantum mechanics by modifying the Heisenberg
Uncertainty Principle (HUP).
\section{Compactness and Quantum Uncertainties}
\setcounter{equation}{0}
\label{S:Cgup}
We describe a compact object by means of a quantum wavefunction $\psi=\psi(x_i)$
with Cartesian coordinates $x_i=(x,y,z)$ in its centre-of-mass reference frame, so that
\be
\bra{\psi}\hat x_i\ket{\psi}
\equiv
\expec{\hat x_i}
=
0
=
\expec{\hat p_i}
\ .
\ee
The restriction to the centre-of-mass reference frame should remove unphysical deviations
from well-established classical dynamics~\cite{Casadio:2020rsj}, an issue known as the
``soccer ball problem''~\cite{Hossenfelder:2012jw}.
We then have the uncertainties 
\be
\Delta x_i^2
\equiv
\expec{\hat x_i^2}-\expec{\hat x^i}^2
=
\expec{\hat x_i^2}
\ee
and
\be
\Delta p_i^2
\equiv
\expec{\hat p_i^2}-\expec{\hat p_i}^2
=
\expec{\hat p_i^2}
\ .
\ee
Furthermore, we assume spherical symmetry, which implies 
\be
\Delta x_i^2
=
\frac{R^2}{3}
\ ,
\label{dxGUP_0}
\ee
and the distribution of momentum uncertainties is isotropic,
\be
\Delta p_x^2 = \Delta p_y^2 = \Delta p_z^2
=
\frac{M^2}{3}
\ ,
\label{dpGUP}
\ee
where the mass of the object is given by the dispersion relation $M^2\simeq p_x^2+p_y^2+p_z^2$. 
\par
The (dimensionless) compactness of a star or black hole is defined as the
ratio~\footnote{From now on, we mostly use units with $\hbar=\lpl\mpl$, and $\gn=\lpl/\mpl$.}
\be
\mathcal{C}
=
\frac{\gn\,M}{R}
=
\frac{\lpl\,M}{\mpl\,R}
\ ,
\label{compactness}
\ee
where now $M$ and $R$ are obtained from Eqs.~\eqref{dpGUP} and \eqref{dxGUP_0}.  
It is important to remark that we employed Cartesian coordinates in the above argument
because all of the operators $\hat x_i$ and $\hat p_i=-i\,\hbar\,\partial_i$ are Hermitian
[on square integrable functions $\psi=\psi(x_i)$ of the real numbers].
Although spherical coordinates would seem a more natural choice to describe spherically
symmetric systems, the operator $\hat p_r=-i\,\hbar\,\partial_r$ is not Hermitian,
and it would not be clear how to implement uncertainty relations to bound the compactness
in that frame. 
\section{Upper bound from GUP}
\setcounter{equation}{0}
\label{S:GUP}
The simple GUP is given by Eq.~\eqref{gup0}, which we rewrite as
\be
\Delta x\,\Delta p
\gtrsim
\lpl\,\mpl
\left(1 + \alpha\, \frac{ \Delta p^2}{\mpl^2}\right)
\ ,
\label{GUP}
\ee
and assume that $\alpha$ is positive.
In the extremal case, in which the equality holds, there are two solutions for $\Delta p$,
to wit
\be
\Delta p_\pm
= 
\frac{\mpl\,\Delta x}{2\,\alpha\,\lpl}
\left( 1 \pm \sqrt{1-4\,\alpha\, \frac{ \lpl^2}{\Delta x^2}} \right)
\ .
\label{Dp+-}
\ee
Note that $\Delta p_-$ reproduces the HUP in the limit $\alpha \to 0^+$ (or for $\Delta x\gg \lpl$).
This restriction will be taken into account to describe objects with size $R\sim\Delta x\gg\lpl$. 
\par
\begin{figure}[t]
\centering
\includegraphics[width=0.8\textwidth]{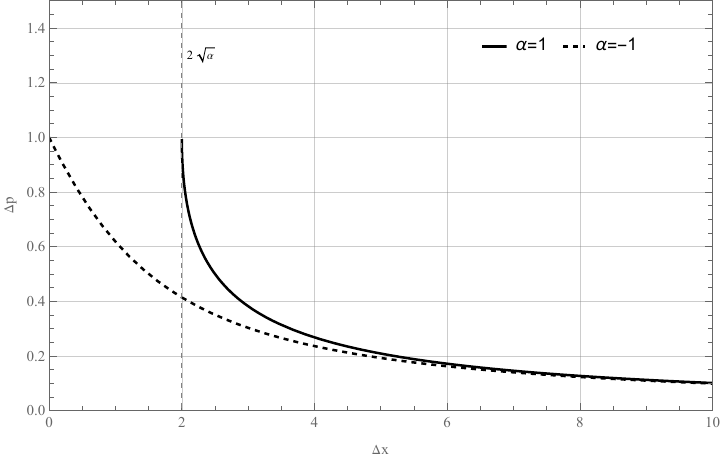}
\caption{$\Delta p=\Delta p_-$ from Eq.~\eqref{Dp+-}, with $\Delta x$ in units of $\lpl$ and $\Delta p$ in units of $\mpl$.}
\label{fig:dxdpgup}
\end{figure}

The above Eq.~\eqref{Dp+-} for $\alpha>0$ yields complex momenta for $\Delta x$ smaller than a minimum length
determined by
\be
\sqrt{1-4\,\alpha\, \frac{ \lpl^2}{\Delta x^2}}=0
\qquad
\Rightarrow
\qquad
\Delta x
=
2\, \sqrt{\alpha}\,\lpl
\equiv
\ell_{\rm GUP}
\ ,
\label{Lgup}
\ee 
which is also called the remnant limit in the thermodynamic community~\cite{Adler_2,Cimidiker:2023kle}.
This limit for the physical solution $\Delta p=\Delta p_-$ from Eq.~\eqref{Dp+-} is shown in Fig.~\ref{fig:dxdpgup}
for positive values of $\alpha$.
We recall that negative values of $\alpha$ do not imply the existence a minimum length scale.
However, this case still results in objects with exotic behaviour such as an infinite lifetime~\cite{Ong:2018syk}.
\par
The inequality in Eq.~\eqref{GUP} can be solved in terms of $\Delta x$,
\be
\Delta x \gtrsim \frac{\lpl\,\mpl}{\Delta p} \left(1 + \alpha \frac{ \Delta p^2}{\mpl^2}\right)
\ ,
\label{dxGUP}
\ee
and replaced in the definition of the compactness from Eq.~\eqref{compactness}, to obtain 
\be
\mathcal{C}
\lesssim
\frac{M^2}{3\,\mpl^2 + \alpha\,M^2} 
\simeq
\frac{1}{\alpha}
\ ,
\label{GUP_limit}
\ee
where the approximate expression is for objects with mass $M\gg\mpl$.
Assuming $\alpha\sim 1$, the conclusion is that the GUP in Eq.~\eqref{GUP} implies an upper
bound for the compactness of order one. The unity of $\alpha$ is further reinforced by the results
in Refs.~\cite{Amati:1988tn,Gross:1987kza,Konishi:1989wk,Capozziello:1999wx}.
\par
It is important to recall that black holes are characterised by $\mathcal{C}>1/2$, and Eq.~\eqref{GUP_limit} 
therefore allows for their formation only provided $\alpha\lesssim 2$,
otherwise the gravitational collapse would lead to horizonless exotic compact objects~\cite{Cardoso:2019rvt}.
On the other hand, the limit $\alpha \to 0$, for fixed $M$, yields
\be
\mathcal{C}
\lesssim
\frac{M^2}{3\,\mpl^2}
\ ,
\label{HUP_limit}
\ee
which implies that the usual HUP does not place any significant bound on the compactness 
of astrophysical compact objects and formation of black holes with $M\gg\mpl$.~\footnote{It is
somewhat intriguing that the r.h.s.~of Eq.~\eqref{HUP_limit} bares a qualitative resemblance with
Bekenstein's quantisation of the black hole area~\cite{bekenstein}.} 
\section{Bounds from GEUP}
\setcounter{equation}{0}
\label{S:GEUP}
We rewrite the GEUP in Eq.~\eqref{geup0} as
\be
\Delta x\,\Delta p
\gtrsim
\lpl\,\mpl \left(1 + \alpha\, \frac{ \Delta p^2}{\mpl^2} + \beta\, \frac{\Delta x^2}{\lpl^2}\right)
\ .
\label{GEUP}
\ee
The additional term proportional to the dimensionless constant $\beta$
is phenomenologically linked to a non-zero cosmological constant~\cite{Bambi:2007ty},
which suggests the existence of a Hawking-Page-like phase transition.
From a thermodynamical perspective, it is also noteworthy that the limit $\alpha \to 0$ of the GEUP
[that is, the EUP in Eq.~\eqref{eup0}] exhibits a functional form identical to that
of the R{\'e}nyi entropy~\cite{Moradpour:2019yiq, Cimdiker:2022ics}.
\par
Again, in the extremal case there are two solutions 
\be
\Delta p_\pm
= 
\frac{\Delta x\, \mpl}{2\,\alpha\,\lpl} \left( 1 \pm \sqrt{1-4\,\alpha\,\beta-4\,\alpha\, \frac{ \lpl^2}{\Delta x^2}} \right)
\ ,
\label{delpGEUP}
\ee
where the solution $\Delta p_-$ again reduces to the HUP for $\alpha \to 0^+$ and any finite value of $\beta$.
Like previously, for positive $\alpha$, we obtain a minimum measurable length determined by
\begin{equation}
\sqrt{1-4\, \alpha\, \beta - 4\,\alpha\, \frac{ \lpl^2}{\Delta x^2} }=0 
\qquad
\Rightarrow
\qquad
\Delta x
=
\frac{2\,\lpl\, \sqrt{\alpha }}{\sqrt{1-4\, \alpha \, \beta }}
\equiv
\ell_{\rm GEUP}
\ ,
\end{equation}
which is real for $4\,\alpha\,\beta<1$.
This is the GEUP remnant limit, and the GUP case~\eqref{Lgup} is recovered for $\beta \to 0$.
Note that the parameter space described above has already been extensively explored in the (near) quantum
regime in Ref.~\cite{bcjm25}.
\par
Both solutions $\Delta p=\Delta p_\pm$ in Eq.~\eqref{delpGEUP} display a sort of {\em phase transition},
where the slope changes sign at
\be
\Delta x_{\rm pt}
=
\frac{\lpl}{\sqrt{\beta -4\, \alpha\,  \beta ^2}}
\ ,
\ee
again provided $\beta>0$ and $4\,\alpha\,\beta<1$.
In the limit $\alpha \to 0$, we obtain $\Delta x_{\rm pt}=\lpl/\sqrt{\beta}$, while $\Delta x_{\rm pt}$
diverges for $\beta \to 0$, which is consistent with the GUP case of Section~\ref{S:GUP}. 
\par
\begin{figure}[t]
\centering
 \includegraphics[width=0.8\textwidth]{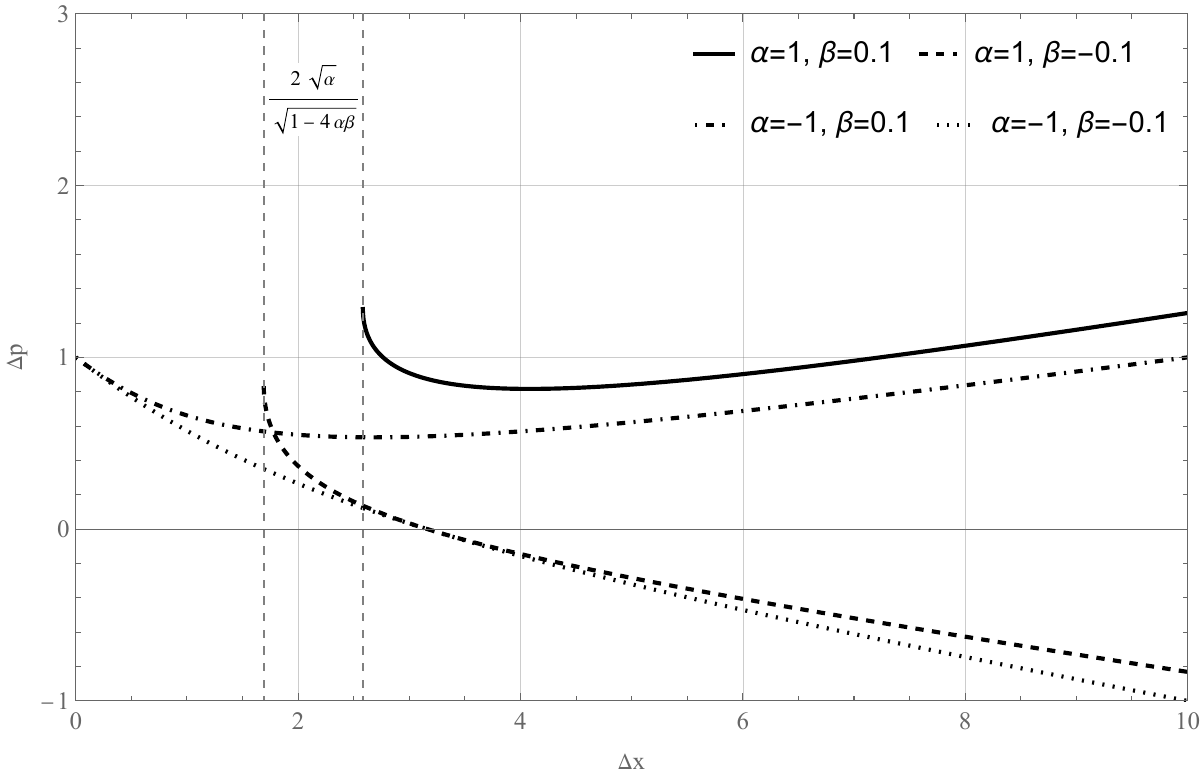}
 \caption{$\Delta p=\Delta p_- $ from Eq.~\eqref{delpGEUP}, with $\Delta x$ in units of $\lpl$ and $\Delta p$ in units of $\mpl$.}
 \label{fig:dxdpgeup}
\end{figure}

In Fig.~\ref{fig:dxdpgeup}, the minimum length values are highlighted for the solution $\Delta p=\Delta p_-$.
We see that positive values of $\alpha$ lead to a minimum length, whereas positive values of $\beta$
yield a phase transition.
Negative values of $\beta$, on the other hand, lead to unphysical negative momentum uncertainties.
\par
Solving the inequality in Eq.~\eqref{GEUP} for $\Delta x$, we obtain the two constraints
\be
\frac{\lpl\,\Delta p}{2\, \beta\,\mpl} \left( 1 - \sqrt{1 - 4\, \alpha\, \beta - 4\, \beta\, \frac{\mpl^2}{\Delta p^2}} \right)
\lesssim
\Delta x
\lesssim
\frac{\lpl\,\Delta p}{2\, \beta\,\mpl} \left( 1 + \sqrt{1 - 4\, \alpha\, \beta - 4\, \beta\, \frac{\mpl^2}{\Delta p^2}} \right)
\ ,
\label{DxGEUP}
\ee
with the solution $\Delta p_-$ yielding the correct HUP limit this time for $\beta \to 0$.
The corresponding compactness range is then given by 
\be
\frac{2\, \beta}{1 + \sqrt{1 - 4\, \alpha\, \beta - 12\, \beta\,{\mpl^2}/{M^2}}} 
\lesssim 
\mathcal{C}
\lesssim 
\frac{2\, \beta}{1 - \sqrt{1 - 4 \,\alpha\, \beta - 12 \,\beta \,{\mpl^2}/{M^2}}}
\ ,
\label{GEUP_limit}
\ee
which, in the large $M$ limit, reduces to
\be
\frac{1-\sqrt{1-4 \,\alpha\, \beta}}{2\, \alpha}
\lesssim
\mathcal{C}
\lesssim 
\frac{1+\sqrt{1-4 \,\alpha\, \beta}}{2\, \alpha}
\ ,
\label{GEUP_limit_as}
\ee
where we again assumed $\beta>0$ and $0 \lesssim 4\,\alpha\,\beta<1$.
Note that the GUP bound~\eqref{GUP_limit} is recovered for $\beta\to 0$, whereas the limit 
$\alpha\to 0$ for finite $\beta>0$ yields a lower bound $\mathcal{C}\gtrsim \beta$.
\par
\begin{figure}[hbt!] 
\centering
\centering
\includegraphics[width=0.8\textwidth]{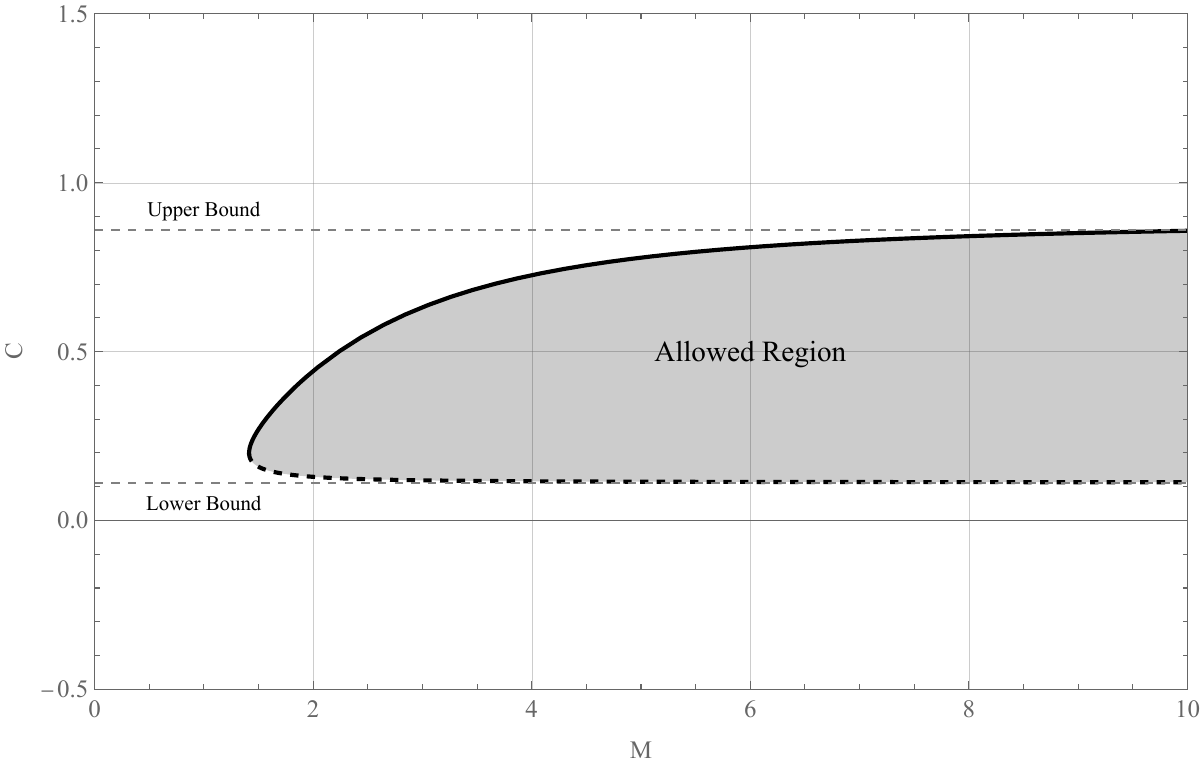}
\caption{Compactness as function of the mass $M$ (in units of $\mpl$).
Shaded region represents values allowed by the inequality~\eqref{GEUP_limit} with $\alpha=1$ and $\beta=0.1$.}
\label{fig:cpos}
\end{figure}
\begin{figure}[hbt!] 
\centering
\includegraphics[width=0.8\textwidth]{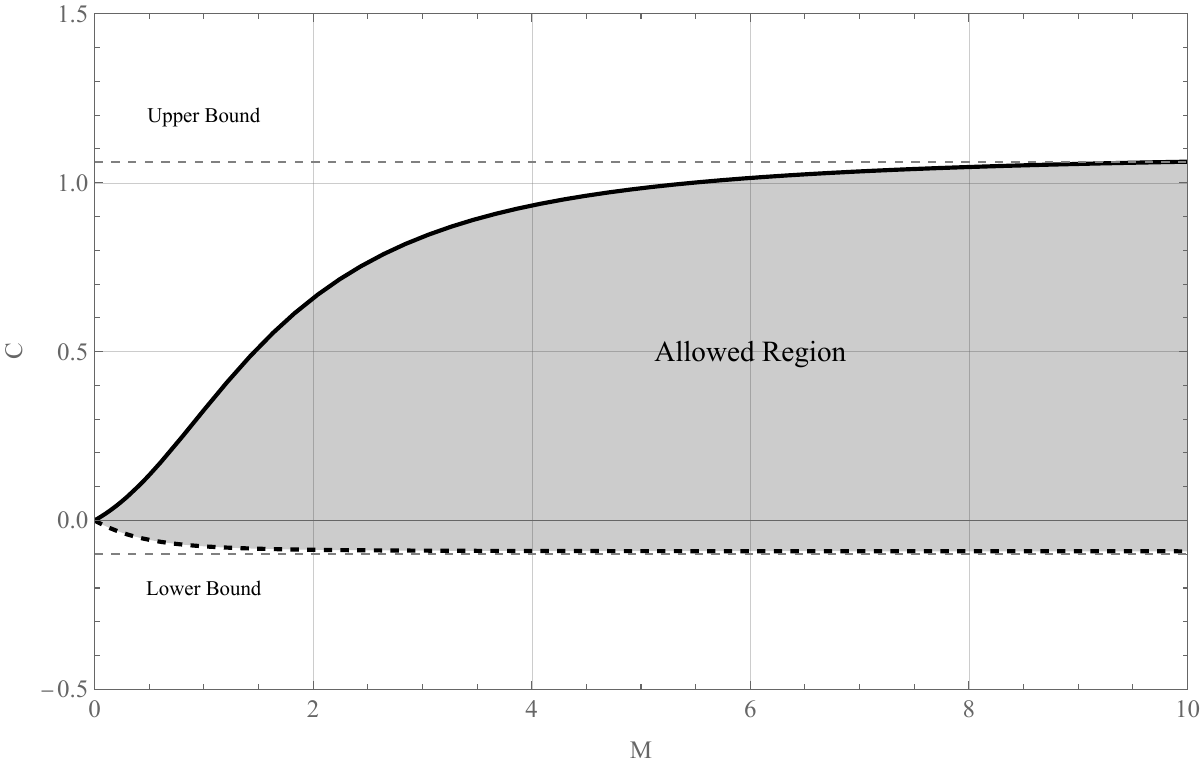}
        \caption{Compactness as function of the mass $M$ (in units of $\mpl$).
Shaded region represents values allowed by the inequality~\eqref{GEUP_limit} with $\alpha=1$ and $\beta=-0.1$.}
        \label{fig:cneg}
\end{figure}

In Fig.~\ref{fig:cpos}, we see that the compactness is asymptotically bounded from below
and above by the values corresponding to inequality~\eqref{GUP_limit}.
Negative values of $\beta$ also yield a valid upper bound displayed in Fig.~\ref{fig:cneg}
(of course, the lower bound is not physically significant for negative $\mathcal C$).
%
\section{Discussion and conclusions}
\label{S:conclusions}
\setcounter{equation}{0}
In this work we have obtained bounds on the compactness of self-gravitating objects
depending on the GUP and GEUP parameters $\alpha$ and $\beta$.
Such parameters are in turn bounded by their effects on experimental observations.
For the GUP, one the strongest constraint $\alpha\lesssim 10^7$ comes from
the study of harmonic oscillators~\cite{Bawaj:2014cda}.
\par
For the EUP, 
bounds are typically placed on the associated length scale $L_*$.
To our knowledge, the strongest of such bounds is obtained from the precession
of S2's orbit and is given by $L_{*} \gtrsim 4\,\sqrt{\bar\beta}\,10^{10}\,$m~\cite{Okcu:2022sio},
which results in 
\begin{equation}
\beta
\lesssim
10^{-90}
\ .
\end{equation}
For such small values of $\beta$, the upper bound from the GEUP in Eq.~\eqref{GEUP_limit_as}
essentially reduces to the GUP bound~\eqref{GUP_limit}.
\par
However, the Event Horizon Telescope collaboration observed the shadow resulting
from the phenomena taking place near the event horizon of the supermassive black
hole candidates in the centre of M87~\cite{EventHorizonTelescope:2019dse} and in the
centre of our Galaxy, Sgr~A*~\cite{EventHorizonTelescope:2022wkp}.
Requiring that black holes do indeed form, one finds that $\mathcal C\sim 1/2$
must be allowed, which then implies that values of $\alpha\sim 2$ should also be acceptable
according to the GUP bound~\eqref{GUP_limit}.
Such cases correspond to a minimum length $\ell_{\rm GUP}\sim 3\,\lpl$,
which is still in qualitative agreement with independent analyses, like those stemming from string 
theory~\cite{Amati:1988tn,Gross:1987kza,Konishi:1989wk,Capozziello:1999wx}.~\footnote{Smaller
values of $\alpha$ are also possible, for instance according to the quantum model
of Ref.~\cite{Casadio:2023ymt} and in the context of bootstrapped Newtonian gravity where objects
with stable cores (of larger compactness) can exist hidden beyond the horizon~\cite{Casadio:2019cux}.}
Moreover, using the same argument, we can place independent bounds on the GEUP 
paramater $\beta$.
Starting from the inequality on the right of Eq.~\eqref{GEUP_limit_as} one can constrain
$\beta$ in terms of the value of $\alpha$.
When both parameters are positive and all the factors contained in the GEUP are real,
the constraint which ensures that the maximum compactness is large enough to allow
for black hole (horizon) formation yields $0 \lesssim \beta\lesssim {1}/{4\,\alpha}$
for $0\lesssim \alpha\lesssim 1$, or $0 \lesssim \beta\lesssim (2-\alpha)/4$ for
$1\lesssim \alpha\lesssim 2$, with values of $\alpha$ outside this range being
excluded.~\footnote{Constraints using the minimum compactness can be imposed in the GEUP case,
but we found those not to be relevant since they depend on an arbitrarily low defined value of this parameter.}

%
%
%
%
%
%
%
%
\section*{Acknowledgments}
R.C.~is partially supported by the INFN grant FLAG.
The work of R.C.~has also been carried out in the framework
of activities of the National Group of Mathematical Physics (GNFM, INdAM)
and COST action CA23115 (RQI). 
I.C.~is supported by the Polish National Science Centre grant No.~DEC-2020/39/O/ST2/02323
and Polish National Agency for Academic Exchange grant No.~BPN/PRE/2022/\-1/00068.
O.M.~is supported by the Romanian Ministry of Research, Innovation and Digitalization
under the Romanian National Core Program LAPLAS VII-contract n.~30N/2023. 
I.C.~and J.M.~thank the Department of Physics and Astronomy at the University of Bologna for
hospitality during the initial stages of this project.
\end{document}